\documentclass[preprint,aps,prd,superscriptaddress,nofootinbib,tightenlines]{revtex4}

\usepackage{amsfonts}
\usepackage{multirow}
\usepackage{mathrsfs}
\usepackage{graphicx}
\usepackage{amsmath}
\usepackage{amssymb}
\usepackage{bm}
\usepackage{bbm}
\usepackage{color}
\usepackage{xcolor}
\usepackage{float}
\usepackage{slashed}
\usepackage{ulem}

\def\HLFz{\mathrm{H}_{\mathrm{LF}}^{(0)}}

\def\QCDtwo{{\mathrm{QCD}_2}}
\def\QCDfour{{\mathrm{QCD}_4}}

\def\Xint#1{\mathchoice
{\XXint\displaystyle\textstyle{#1}}%
{\XXint\textstyle\scriptstyle{#1}}%
{\XXint\scriptstyle\scriptscriptstyle{#1}}%
{\XXint\scriptscriptstyle\scriptscriptstyle{#1}}%
\!\int}
\def\XXint#1#2#3{{\setbox0=\hbox{$#1{#2#3}{\int}$ }
\vcenter{\hbox{$#2#3$ }}\kern-.6\wd0}}

\def\dashint{\Xint-}

\newcommand{\nn}{\nonumber}

\newcommand{\beq}{\begin{equation}}
\newcommand{\eeq}{\end{equation}}
\newcommand{\bqa}{\begin{eqnarray}}
\newcommand{\eqa}{\end{eqnarray}}
\newcommand{\bseq}{\begin{subequations}}
\newcommand{\eseq}{\end{subequations}}

\begin{document}
%\preprint{}
\title{\mbox{}\\[10pt]
Heavy quark fragmentation function in 't Hooft Model}

\author{Yu Jia~\footnote{jiay@ihep.ac.cn}}
\affiliation{Institute of High Energy Physics, Chinese Academy of Sciences, Beijing 100049, China}
\affiliation{School of Physics, University of Chinese Academy of Sciences, Beijing 100049, China}

\author{Zhewen Mo~\footnote{mozw@itp.ac.cn}}
\affiliation{Institute of High Energy Physics, Chinese Academy of Sciences, Beijing 100049, China}
\affiliation{CAS Key Laboratory of Theoretical Physics, Institute of Theoretical Physics,Chinese Academy of Sciences, Beijing 100190, China}

\author{Xiaonu Xiong~\footnote{xnxiong@csu.edu.cn}}
\affiliation{School of Physics, Central South University, Changsha 418003, China}

%%%%%%%%%%%%%%%%%%%%%%%%%%%%%%%%%%%%%%%%%%%%%%%%%%%%%%%%%%%%%%%%%%%%%%%%%%%%%%
\begin{abstract}
    We carry out a comprehensive study of the quark-to-meson fragmentation function in the 't Hooft model, {\it i.e.},
      the two-dimensional Quantum Chromodynamics (QCD) in $N_c\to \infty$ limit, following the operator definition pioneered by Collins and Soper.
      We apply the Hamiltonian approach as well as the diagrammatic approach to construct the functional form of  the quark-to-meson fragmentation function in terms of the meson's
      light-cone wave function.
      For the sake of comparison, we also investigate the heavy quark fragmentation into quarkonium in two-dimensional QCD
      within the framework of the nonrelativistic QCD (NRQCD) factorization, at the lowest order in quark velocity.
      In the heavy quark limit, the quark fragmentation function obtained from the {\it ab initio} method agrees well, both analytically and numerically,
      with that obtained from the NRQCD approach. This agreement might be regarded as a nontrivial justification for the
      validity of both field-theoretical approaches to compute the heavy quark fragmentation function.
    \end{abstract}

    %\pacs{}

    %%%%%%%%%%%%%%%%%%%%%%%%%%%%%%%%%%%%%%%%%%%%%%%%%%%%%%%%%%%%%%%%%%%%%%%%%%%%%%
    \maketitle
    %%%%%%%%%%%%%%%%%%%%%%%%%%%%%%%%%%%%%%%%%%%%%%%%%%%%%%%%%%%%%%%%%%%%%%%%%%%%%%

    \section{Introduction}
    \label{intro}

    Quantum Chromodynamics (QCD) is the fundamental quantum field theory describing the strong interaction.
    Although tremendous progress in understanding all aspects of strong interaction
    has been made over the past half century, some extraordinary features of QCD still remain as a mystery, such as the color
    confinement mechanism, and how the colored quarks and gluons transition into the color-singlet hadrons. Fragmentation
    functions (FF) are universal nonperturbative functions that characterize the probability for a high-energy parton
    hadronizing into an identified hadron carrying a definite momentum fraction,
    which are closely related to the aforementioned profound myths in QCD.

    According to the celebrated QCD factorization theorem, the FFs are essential theoretical input to describe the identified hadron
    production at large $p_T$ in high energy collision experiments. Being universal nonperturbative functions, the FFs have been directly extracted
    from the $e^+e^-$ and semi-inclusive deep inelastic scattering experiments, or parameterized by some phenomenological models~\cite{Kretzer:2000yf,Kretzer:2001pz,deFlorian:2007ekg,Bertone:2017tyb}.
    Needless to say, it is highly desirable to deduce the FFs from a rigorous field-theoretical perspective.

    So far lattice QCD proves to be the only systematic and reliable nonperturbative approach to
    unravel the hadron internal structure.
    Thanks to the advent of large momentum effective theory (LaMET){~\cite{Ji:2013dva,Ji:2014gla,Ji:2020ect}}, we have witnessed tremendous progress
    in the last decade in computing a hadron's parton distribution functions (PDFs) and light-cone distribution ampliutdes (LCDA)
    directly in the $x$ space through the lattice Monte Carlo simulation {(for a recent review, see~\cite{Lin:2023kxn})}. Unfortunately, it is impossible for lattice QCD to
    compute the FFs, even in the framework of LaMET, because the summation over the colored out-state which emerges from
    the definition of fragmentation functions poses a insurmountable obstruction for lattice simulation.

    Since the first-principle determination of the FFs in QCD looks impractical in the foreseeable future,
    we may seek to learn some useful lessons from the toy models of QCD. The 't Hooft model, {\it i.e.},
    the 1+1 dimensional QCD ($\QCDtwo$) in the $N_c\to \infty$ limit, is a solvable model, which resembles the realistic
    ${\rm QCD}_{4}$  in several aspects, such as color confinement, Regge trajectory, quark condensate and even (naive)
    asymptotic freedom. In the past, the meson's PDFs, generalized PDFs, as well as the quasi-PDFs{~\cite{Burkardt:2000uu,Jia:2018qee,Ji:2018waw,Ma:2021yqx}} have
    been comprehensively studied within the 't Hooft model.
    It is the goal of this work to investigate the fragmentation function in this toy model of QCD.
    Due to the lack of the transverse degrees of freedom in, the gluon cannot be a dynamical parton, therefore we
    restrict ourselves in considering only the quark-to-meson fragmentation function.

    In the late 70s, Einhorn~\cite{Einhorn:1976ev} and Ellis~\cite{Ellis:1977mw} have already discussed how to extract the quark fragmentation
    function through the fictitious high-energy process $e^-e^+\to H+X$ in $\QCDtwo$. Nevertheless, there are some serious shortcomings in this
    ``experimental" approach. Firstly, whether the factorization theorem in two spacetime dimension
    is valid or not in $\QCDtwo$ is unclear. Secondly, their expressions are rather complicated,
    which involve three mesonic light-cone wave functions, two of which are affiliated with the infinitely high excited mesonic states.
    The exceedingly oscillatory behavior of these wave functions obstruct the numerical study of the FFs.
    Therefore, in this work we start from the operator definition of the FF pioneered by Collins and Soper, and we are able to
    express the quark-to-meson FF in terms of a single light-cone wave functions (LCWFs)  associated with the identified meson.
    The virtue of our approach is that our expression is quite compact,
    and amenable to numerical studies.

    Of special interest is the heavy quark fragmentation function. Thanks to the asymptotic freedom, the NRQCD factorization approach allows one to
    further refactorize the heavy quark fragmentation into the heavy quarkonium
    into the product of the short-distance coefficients and universal long-distance NRQCD matrix element.
    In this work, we also compute the quark-to-quarkonium fragmentation function using NRQCD approach in 't Hooft model.

    It is reassuring that, in the heavy quark limit,
    the quark fragmentation function obtained from the {\it ab initio} method is well consistent with the NRQCD prediction.
    This agreement can be considered as strong evidence that our understanding of fragmentation functions is correct, at least in
    the two-dimensional QCD.

    The rest of the paper is organized as follows:
    %-------------------------------
    In Sec.~\ref{sec:def:FF}, we write down the Collins-Soper's definition of quark fragmentation function,
    which can be readily carried over to 1+1 spacetime dimension.
    %-------------------------
    In Sec.~\ref{sec:QCD2} we briefly review the essential ingredients of the 't Hooft model.
    %-------------------------
     In Sec.~\ref{sec:operator} we discuss how to deduce the analytical expression for the quark fragmentation function
     using the operator approach.
    %-------------------------
    In Sec.~\ref{sec:nrqcd} we present the derivation of the heavy quark fragmentation function
    using NRQCD factorization approach, accurate at lowest order in velocity expansion.
    %-------------------------
    In Sec.~\ref{comparison:two:approaches}, we analytically prove that, in the heavy quark limit,
    the heavy quark fragmentation function obtained from the {\it ab initio} approach and NRQCD approach
    are compatible with each other.
%-------------------------
In Sec.~\ref{sec:numerical}, we present numerical results of the FFs in 't Hooft model, and also compare the
{\it ab initio} and NRQCD results.
%-------------------------
Finally in Sec.~\ref{sec:sum} we present a summary and outlook.
%-------------------------
In Appendix we also present an alternative derivation of the quark fragmentation function using Feynman diagrammatical approach.
%-------------------------

\section{Definition of quark-to-meson fragmentation function}\label{sec:def:FF}

Fragmentation functions are opposite to the PDFs, which encapsulate the probability of finding an identified hadron
with a certain fractional light-cone momentum with respect to the parent parton. The FFs are important to describe the
inclusive production of an identified hadron. For instance, to describe the
inclusive production of the the $B^-$ meson from $Z^0$ boson decay, the QCD factorization theorem indicates the leading
contribution to be
%-------------------------
\begin{align}
%-------------------------
& d\Gamma(Z^0\to B^-(E)+X) =  \int dz\,\hat{\Gamma}(Z^0\to b(E/z) \bar{b})) \, D_{b\to B^-}(z)+\cdots,
%-------------------------
\end{align}
%-------------------------
where $E$ denotes the energy of the $B^-$ meson, and $z$ denotes the light-cone momentum fraction carried by $B^-$ with respect to the $b$ quark.
$D_{b\to B^-}(z)$ denotes the $b$-to-$B^-$ fragmentation function, and $\hat{\Gamma}$ denotes the partonic decay rate for $Z^0\to b\bar{b}$.

The quark fragmentation function also admits a rigorous operator definition,
first introduced by Collins and Soper in 1981~\cite{Collins:1981uw}
%-------------------------
\begin{align}
%-------------------------
    \notag D_{q\rightarrow H}(z)=&\sum_{X}\frac{z^{d-3}}{4\pi}\int d\xi^- e^{-iz^{-1}P^+\xi^-}\frac{1}{N_c}\mathrm{Tr_{color}}\mathrm{Tr_{Dirac}}\\
    \notag &\times \gamma^+\left\langle 0\left| \tilde{\mathrm{T}}\, \mathcal{W}\left[\infty,0\right]\psi(0)\right|H+X\right\rangle \\
    &\times\left\langle X+H\left|\mathrm{T}\,\bar{\psi}\left(\xi^-\right)\mathcal{W}\left[\xi^-,\infty\right]\right|0\right\rangle,
%-------------------------
\label{eq:csdef}
%-------------------------
    \end{align}
%-------------------------
where the light-cone coordinates $x^\pm={(x^0 \pm x^z)}/{\sqrt{2}}$ are used. $d$ signifies the spacetime dimension
(In our case, $d$ will be put to 2).
$H$ represents the identified color-singlet hadron, while $X$ denotes the unobserved soft hadrons (however, here $X$ has to be an colored object due to color conservation).
$\mathrm{T}$ and $\tilde{\mathrm{T}}$ represent the time-ordering and anti time-ordering.
$\mathcal{W}[\xi^-,\eta^-]$ is the path-ordered exponential along the light-cone ``$-$'' direction:
%-------------------------
\beq
%-------------------------
\mathcal{W}[\xi^-,\eta^-]=\mathcal{P}\left\{\exp\left[-i\int_{\xi^-}^{\eta^-} d\zeta^- A^+(\zeta^-)\right]\right\},
%-------------------------
\label{eq:W:def}
%-------------------------
\eeq
%-------------------------
whose role is to ensure the gauge invariance of the FF.
If the light-cone gauge $A^{+, a} = 0$ is imposed, this gauge link can be simply dropped.

\section{A Brief review of 't Hooft model: Hamiltonian organized in $1/N_c$ expansion}\label{sec:QCD2}

In this section, we recap some necessary ingredients of 't Hooft model. We will mainly concentrate on the
Hamiltonian approach and bosonization program, which will be used to derive the quark fragmentation function in the next section. For more comprehensive discussions, we refer the interested readers to {Refs.~\cite{Lenz:1991sa,Barbon:1994au,Jia:2018qee}.}

For simplicity, we will consider the $\QCDtwo$ with a single quark flavor:
%---------------------------
\begin{equation}
    \mathcal{L}_\QCDtwo =
    - \frac{1}{4}F^{\mu\nu,a}F^{a}_{\mu\nu}
    + \overline\psi\left(i\slashed{D} -m\right)\psi,
   \label{QCD:lagr}
\end{equation}
%---------------------------
where $\psi$ denotes the 2-component Dirac field for quark, and $m$ signifies the current quark mass.
The color covariant derivative $D_\mu$ is defined as $D_\mu= \partial_\mu-ig_s A_\mu^aT^a$,
with $T^a$ denoting the $SU(N_c)$ generators in the fundamental representation. $g_s$ signifies the dimensionful coupling constant in $\QCDtwo$.
The gluon field strength tensor is given by $F_{\mu\nu}^a \equiv \partial_\mu A_\nu^a-\partial_\nu A_\mu^a+g_sf^{abc}A_\mu^bA_\nu^c$.
The 't Hooft model also needs taking the limit $N_c\rightarrow\infty$
while keeping the 't Hooft coupling $\lambda \equiv \frac{g_s^2N_c}{4\pi}$ fixed.
$\sqrt{2\lambda}$ can be regarded as the characteristic mass scale in the 't Hooft model£¬
similar to $\Lambda_{\rm QCD}$ in the realistic QCD.

Adopting the chiral-Weyl representation for the Dirac $\gamma$ matrices:
%---------------------------
\begin{equation}
    \gamma^0=\sigma_1,\quad \gamma^1=-i\sigma_2,\quad \gamma_5\equiv \gamma^0\gamma^1=\sigma_3,
\nn
\end{equation}
%---------------------------
we can express the Dirac spinor field as
%---------------------------
\begin{equation}
%---------------------------
    \psi = 2^{-{1\over 4}}
    \left(
    \begin{array}{c}
        \psi_R \\ \psi_L
    \end{array}
    \right),
    \label{psidcp}
%---------------------------
\end{equation}
%---------------------------
where $R$, $L$ denote the right-handed and left-handed components.

Substituting \eqref{psidcp} into \eqref{QCD:lagr}, and imposing the light-cone gauge $A^{+, a} = 0$,
one obtains~\cite{tHooft:1974pnl}
%---------------------------
\begin{align}
 \notag &\mathcal{L}_\QCDtwo
    =\frac{1}{2}
    \left(
        \partial_{-} A^{-,a}
    \right)^{2}
    + g_{s}\psi_{R}^{\dagger}A^{-,a}T^{a}\psi_{R}\\
    &+\psi^{\dagger}_{R} i \partial_{+}\psi_{R}
    +\psi^{\dagger}_{L} i \partial_{-}\psi_{L}
    - \frac{m}{\sqrt{2}}
    \left(
        \psi^{\dagger}_{L}\psi_{R}+
        \psi^{\dagger}_{R}\psi_{L}
    \right).
\end{align}
%---------------------------

One immediately solve the equations of motion for $A^{-,a}$ and $\psi_{L}$:
%---------------------------
\bseq
\begin{align}
    &\partial_{-}^{2}A^{-,a}-g_{s}\psi^{\dagger}_{R}T^{a}\psi_{R}=0,
    \\
    & i \partial_{-}\psi_{L}-\frac{m}{\sqrt{2}}\psi_{R}=0.
\end{align}
\eseq
%---------------------------
Since the derivative is with respect to the light-cone position $x^-$ rather than light-cone time $x^+$,
the fields $A^{-,a}$ and $\psi_L$ are no longer propagating degrees of freedom, yet simply regarded as the constraints.
Both of them can be expressed as the canonical variable $\psi_R$ (the ``good'' component)
%---------------------------
\bseq
%---------------------------
\begin{align}
&\psi_L(x^+\!,x^-)\!=\!-i\frac{m}{\sqrt{2}}\int dy^- G^{(1)}(x^--y^-)\psi_R(x^+,y^-),
%---------------------------
\\
%---------------------------
&A^{-,a}(x^+\!,x^-)=g_s\!\!\int\!dy^-\! G^{(2)}(x^-\!\!\!-\!y^-)\psi^\dagger_R(y^-)T^a\psi_R(x^+\!,y^-),
\end{align}
%---------------------------
\eseq
%---------------------------
where $G^{(1)}$ and $G^{(2)}$ are the Green functions of the differential
operators $\partial_{-}$ and $\partial_{-}^{2}$:
%---------------------------
\bseq
%---------------------------
\begin{align}
        G^{(1)}(x^{-}-y^{-})&=i\int
        \frac{dk^{+}}{2\pi}
        \Theta{(|k^+|-\rho)}\frac{e^{-i k^{+}(x^{-}-y^{-})}}{k^{+}},
        \\
        G^{(2)}(x^{-}-y^{-})&=-\int
        \frac{dk^{+}}{2\pi}
        \Theta{(|k^+|-\rho)}
        \frac{e^{-ik^{+}(x^{-}-y^{-})}}{(k^{+})^{2}}.
    \end{align}\label{eq:Greenfunc}
    \eseq
$\Theta$ denotes the Heaviside step function. $\rho$ is an artificial IR cutoff introduced to regularize the IR divergence caused
by exchanging an instantaneous gluon, which must disappear in any physical quantity after taking $\rho\to 0^S+$ limit.

After Legendre transformation, one arrives at the light-front (LF) Hamiltonian~\cite{tHooft:1974pnl}
%---------------------------
\begin{align}
%---------------------------
        \notag {\mathrm{H_{LF}}}=& \int_{x^{+}=\text{const.}}\!\!\!\!\!\!\!\!\!\!\!\!dx^{-}
        \left[
            \frac{m^{2}}{2i}\psi^{\dagger}_{R}(x^{-})
            \!\int dy^{-} G^{(1)}(x^{-}\!-\!y^{-})\psi_{R}(y^{-})
        \right.
%---------------------------
        \nonumber \\
 %---------------------------             &
        & \left.
            - \frac{g_{s}^{2}}{2}\
              \sum_{a} \psi^{\dagger}_{R}(x^{-})T^{a}\psi_{R}(x^{-})
              \int dy^{-} G^{(2)}(x^{-}-y^{-})
              \right.
%---------------------------
              \nonumber\\
%---------------------------
&\left. \times\psi^{\dagger}_{R}(y^{-})T^{a}\psi_{R}(y^{-})
\vphantom{\frac{m^2}{2i}}\right].
%---------------------------
\label{LF:Hamiltonian}
%---------------------------
\end{align}
%---------------------------

At $x^+=0$, the right-handed quark field can be Fourier-expanded as follows:
%---------------------------
\beq
%---------------------------
\psi_R^{i}(x^-)=\int_0^\infty\frac{dk^+}{2\pi}\left[b^i(k^+)e^{-ik^+x^-}+d^{i\dagger}(k^+)e^{ik^+x^-}\right],\label{eq:fieldquant}
%---------------------------
\eeq
%---------------------------
with $i=1,\cdots, N_c$ indicating the color index. The quark(antiquark) annihilation/creation operator $b/b^\dagger$ ($d/d^\dagger$)
obeys the standard anti-commutation relations:
%---------------------------
\begin{equation}
%---------------------------
\{ b^{i\dagger}(k^+), b^j(p^+)\} = 2\pi \delta^{ij} \delta(k^+-p^+),  \qquad  {\{d^{i \dagger}(k^+), d^j(p^+)\} = 2\pi \delta^{ij} \delta(k^+-p^+)},
%---------------------------
\label{eq:bb}
\end{equation}
%---------------------------
and all other unspecified anticommutators simply vanish.

Substituting \label{eq:bb} into the light-front Hamiltonian \eqref{LF:Hamiltonian}, one faces various
bilinear terms composed of quark/antiquark annihilation and
creation operators.  It is convenient to adopt the bosonization technique to expedite the diagonalization of Hamiltonian~\cite{Kikkawa:1980dc,Nakamura:1981zi,Rajeev:1994tr,Dhar:1994ib,Dhar:1994aw,Cavicchi:1993jh,Barbon:1994au,Itakura:1996bk}
by introducing the following bosonic compound operators~\footnote{Note the normalization of the compound operators $B$, $D$ here
differs from what is given in our previous work~\cite{Jia:2018mqi}. The purpose of making this change is to
make the $1/N_c$ expansion of the Hamiltonian manifest.}:
%---------------------------
\bseq
%---------------------------
\begin{align}
%---------------------------
M(k^+,p^+)&=\frac{1}{\sqrt{N_c}}\sum_i d^i(k^+)b^i(p^+),\\
B(k^+,p^+) &= \sum_i b^{i\dagger}({k^+})b^i({p^+}), \\
D(k^+,p^+) &= \sum_i d^{i\dagger}({k^+})d^i({p^+}),
\end{align}
\label{Def:bosonic:compound}
%---------------------------
\eseq
%---------------------------

The LF Hamiltonian can be organized in powers of $1/N_c$:
%---------------------------
\begin{align}
\mathrm{H_{LF}} = \mathrm{H_{LF,vac}}+ \mathrm{H}^{(0)}_{\rm LF} + \mathrm{V}.
%---------------------------
\label{eq:HLF_full}
%---------------------------
\end{align}
%---------------------------

The leading $\mathcal{O}(N_c)$ term corresponds the vacuum energy:
%---------------------------
\begin{align}
%---------------------------
\mathrm{H_{LF,vac}} = N_c \int\frac{dx^-}{2\pi}\left(\frac{\lambda}{2}+\frac{\lambda-m^2}{2}\int_{\rho}^\infty \frac{dk^+}{k^+}\right),
%---------------------------
\end{align}
%---------------------------
which is badly UV and IR divergent~\cite{Lenz:1991sa}.
However, since it is {proportional to} the unit operator and does not have any impact on the FFs, so we will simply drop $\mathrm{H_{LF,vac}}$ henceforth.

The $\mathcal{O}(N_c^0)$ piece in \eqref{eq:HLF_full} reads
%---------------------------
\begin{align}
%---------------------------
\notag &\HLFz =\int_{\rho}^{+\infty}\frac{dk^{+}}{2\pi}\left(\frac{m^2-2\lambda}{2}\frac{1}{k^{+}}+\frac{\lambda}{\rho}\right)\left[B({k^+,k^+})\right.\\\
\notag &\left.+D({k^+,k^+})\right]\!-\!\frac{\lambda}{8\pi^{2}}\int_{0}^{\infty}d k_{1}^{+}d k_{2}^{+}d k_{3}^{+}d k_{4}^{+}\Theta(|k_{1}^{+}\!-\!k_{2}^{+}|\!-\!\rho)\\ &\quad\times\frac{\delta(k_{1}^{+}\!-\!k_{2}^{+}\!-\!k_{3}^{+}\!+\!k_{4}^{+})}{(k_{1}^{+}-k_{2}^{+})^{2}}{M}^{\dagger}(k_{1}^{+},k_{4}^{+}){M}(k_{2}^{+},k_{3}^{+}).
%---------------------------
\label{eq:H0}
%---------------------------
\end{align}
%---------------------------

The last operator $\mathrm{V}$ in the LF Hamiltonian in \eqref{eq:HLF_full} scales as ${\cal O}(1/\sqrt{N_c})$.
For our purpose of computing quark fragmentation function, it is sufficient to know
%---------------------------
\begin{align}
%---------------------------
\mathrm{V}=&-\frac{\lambda}{4\pi^{2}\sqrt{N_{\text{c}}}}\int_{0}^{\infty}\!\!dk^+_{1}dk^+_{2}dk^+_{3}dk^+_{4} \frac{\delta\left(k^+_{1}\!+\!k^+_{2}
\!+\!k^+_{3}\!-\!k^+_{4}\right)}{\left(k^+_{1}+k^+_{3}\right)^{2}}
%---------------------------
\nn\\
%---------------------------
&\times M^{\dagger}(k^+_1,k^+_2)B(k^+_3,k^+_4)+\text{h.c.}+\cdots.
\label{eq:V}
%---------------------------
\end{align}
%---------------------------
These operators induces a process where a quark transitions into a meson and a quark. We have suppressed all other irrelevant operators in $\mathrm{V}$,
such as $M^\dagger D$, $D M$, which instead contribute to the antiquark fragmentation function.

The complete expression of the bosonized LF Hamiltonian can be found in Ref.~\cite{Jia:2018qee}.

Those bosonic compound operators in \eqref{Def:bosonic:compound} are not independent.
Actually $B$, $D$ can be expressed as the convolution between $M$ and $M^\dagger$:
%---------------------------
\bseq
%---------------------------
\begin{align}
%---------------------------
B(k^+,p^+)&=\int_0^\infty \frac{dq^+}{2\pi} M^{\dagger}(q^+,k^+)M(q^+,p^+),
\\
D(k^+,p^+)&=\int_0^\infty \frac{dq^+}{2\pi} M^{\dagger}(k^+,q^+)M(p^+,q^+).
%---------------------------
\end{align}
%---------------------------
\label{eq:BD2M}
%---------------------------
\eseq
%---------------------------
The underlying reason is that, in a confining theory like $\QCDtwo$,
an isolated quark or antiquark can not be created or annihilated from vacuum,
rather only color-singlet quark-antiquark pair can be
created or annihilated~\cite{Kalashnikova:2001df}.

Substituting \eqref{eq:BD2M} into \eqref{eq:HLF_full}, the LF Hamiltonian can be expressed solely in terms of $M$ and $M^\dagger$.
To facilitate  the diagonalization of $\HLFz$, it is convenient to introduce a new set of mesonic annihilation/creation operators $m_n/m_n^\dagger$,
which are related to $M$ and $M^\dagger$ through
%---------------------------
\bseq
%---------------------------
\begin{align}
%---------------------------
&m_n(P^+)=\sqrt{\frac{P^+}{2\pi}}\int_0^1 dx\phi_n(x)M((1-x)P^+,xP^+),
\\
%---------------------------\\
&M((1-x)P^+,xP^+)=\sqrt{\frac{2\pi}{P^+}}\sum_{n=0}^{\infty}\phi_n(x)m_n(P^+),
%---------------------------
\end{align}
%---------------------------
\label{eq:m2M}
%---------------------------
\eseq
%---------------------------
where the coefficient functions $\phi_n(x)$ later will be identified with the 't Hooft light-cone wave function (LCWF) of the $n$-th excited mesonic state, with
$x\in (0,1)$ denote the light-cone momentum fraction carried by the quark inside the meson.

If the mesonic annihilation and creation operators are required to obey the standard commutation relation:
%---------------------------
\beq
%---------------------------
\left[m_n(P_1^+),m^\dagger_r(P_2^+)\right]=2\pi\delta_{nr}\delta(P_1^+-P_2^+),
%---------------------------
%---------------------------
\eeq
%---------------------------
the 't Hooft wave functions must obey the following orthogonality and completeness conditions:
%---------------------------
\bseq
%---------------------------
\begin{align}
&\int_0^1 dx \phi_n(x)\phi_r(x)=\delta_{nr},
%---------------------------
\\
%---------------------------
&\sum_n \phi_n(x)\phi_n(y)=\delta(x-y).
%---------------------------
\end{align}
%---------------------------
\eseq
%---------------------------

Substituting eq.~\eqref{eq:BD2M},~\eqref{eq:m2M} into the LF Hamiltonian eq.~\eqref{eq:H0},
our goal is to have a simple diagonalized form of $\HLFz$, which describe
an infinite towers of non-interacting mesons:
%---------------------------
\beq
%---------------------------
\HLFz= \sum_n \int_0^\infty\frac{dP^+}{2\pi}P_n^-m_n^\dagger(P^+)m_n(P^+),
%---------------------------
\eeq
%---------------------------
where $P_n^-=M_n^2/(2P^+)$ denotes the light-cone energy of the $n$-th excited state meson with light-cone momentum $P^+$.

To achieve this simple form, one must enforce all the off-diagonal terms in $\HLFz$ cancel,
which result in the following
constraints on the coefficient functions $\phi_n(x)$:
%---------------------------
\beq
%---------------------------
\left(\frac{m^2\!-\!2\lambda}{x}\!+\!\frac{m^2\!-\!2\lambda}{1-x}\!-\!M_n^2\right)\phi_n(x) = 2\lambda
\dashint_0^1 \frac{dy}{(x-y)^2}\phi_n(y).
%---------------------------
\label{eq:tHooft_eq}
%---------------------------
\eeq
%---------------------------
This is nothing but the celebrated 't Hooft equation in $\QCDtwo$, the bound-state equation for the $n$-th excited mesonic state on the light-cone
in the 't Hooft model.
Note that the dashed integral $\int\!\!\!\!\!-$ in \eqref{eq:tHooft_eq} indicates a standard principle-value prescription, which originates from taking the $\rho\to 0^+$ limit,
whose role is to tame the IR divergence as $y\to x$.

The single mesonic state can be constructed as
 %---------------------------
\begin{align}
%---------------------------
\left|P^+, P_n^- \right\rangle = \sqrt{2P^+}m_n^\dagger(P^+)\left|0\right\rangle.
%---------------------------
\label{singlet:meson:def}
\end{align}
%---------------------------

To compute the fragmentation function, we must include the color-suppressed operator ${\rm V}$ in the LF Hamiltonian
\eqref{eq:HLF_full}. It will be treated as a perturbation in the context of the $1/N_c$ expansion.

\section{Rigorous derivation of the quark fragmentation function in the leading order in $1/N_c$}
\label{sec:operator}

With all the necessary ingredients at hand, we are ready to conduct an {\it ab initio} derivation of
the quark-to-meson fragmentation function in the 't Hooft model, accurate to the lowest order in $1/N_c$.

For definiteness, we identify the daughter hadron $H$ in the fragmentation function with the $n$-th excited meson in the 't Hooft model.
The asymptotic out state in the far future, $\vert H+X\rangle$, which arises in the definition of the fragmentation function in \eqref{eq:csdef},
needs some careful treatment since it is a colored object.

As shown in \eqref{singlet:meson:def}, the identified color-singlet mesonic state
$\left|H\right\rangle$ is certainly the eigen-state of the leading-color LF Hamitonian:
%---------------------------
\beq
%---------------------------
\HLFz\left|H(P^+)\right\rangle =\frac{M_H^2}{2P^+} |H(P^+)\rangle,
%---------------------------
\eeq
%---------------------------
with $M_H$ signifying the mass of the meson $H$.

But how to treat $|X\rangle$? Color confinement implies that only color-singlet states are allowed to emerge
in the physical spectrum. Clearly $X$ is colored object. Due to color conservation, it must furnish the fundamental
representation of the $SU(N_c)$ group. For simplicity, let us approximate $X$ to be a single dressed quark, which is
indeed true at the leading order in $1/N_c$ expansion.
Curiously, inspecting \eqref{eq:H0},
using the commutation relation
%---------------------------
\beq
%---------------------------
\left[\HLFz ,b^{i\dagger}(p^+)\right] = \left( \frac{m^2\!-\!2\lambda}{2}\frac{1}{p^{+}}\!+\!\frac{\lambda}{\rho}\right)b^{i\dagger}(p^+),
%---------------------------
\label{commutator:HLF:b+}
%---------------------------
\eeq
%---------------------------
one can show that a colored quark is still the eigenstate of the LF Hamiltonian $\HLFz$:
%---------------------------
\bqa
%---------------------------
\HLFz\left|X(p^+)\right\rangle && =\HLFz \sqrt{2p^+} b^{i\dagger}\left(p^+\right)\vert0\rangle
%---------------------------
\nn \\
%---------------------------
&& =\left(\frac{m^2-2\lambda}{2}\frac{1}{p^{+}}+\frac{\lambda}{\rho}\right)\vert X(p^+)\rangle,
%---------------------------
\label{eq:H0_on_quark}
%---------------------------
\eqa
%---------------------------
though the corresponding eigen-energy is gauge-dependent and badly diverges in the limit $\rho\to 0^+$.
Recall all our discussion is confined in the physical light-cone gauge.
This strange feature does not contradict the color confinement, since it needs cost an infinite amount of
energy to excite a single quark state, which is beyond the scope of physical spectrum.

Therefore,  $\vert H+X \rangle$ is also the eigenstate of $\HLFz$:
%---------------------------
\begin{align}
%---------------------------
\notag &\HLFz \vert H(P^+)\!+\!X(p^+)\rangle = \HLFz \sqrt{2p^+} b^{i\dagger}(p^+) \vert H(P^+) \rangle\!= P^-_{H+X} \vert H(P^+)\!+\!X(p^+)\rangle
%---------------------------
\\
%---------------------------
&=  \left(\frac{M_H^2}{2P^+}\!+\!\frac{m^2\!-\!2\lambda}{2}\frac{1}{p^{+}}\!+\!\frac{\lambda}{\rho}\right) \vert H(P^+)\!+\!X(p^+)\rangle,\label{eq:PHX}
%---------------------------
\end{align}
%---------------------------

Using \eqref{singlet:meson:def}, the state $\left|H+X\right\rangle$ can be reexpressed in terms of meson creation operator
acting on the colored soft spectator:
%---------------------------
\beq
%---------------------------
\vert H(P^+)\!+\!X(p^+)\rangle
=\sqrt{2P^{+}}m_n^{\dagger}\left(P^{+}\right)\vert X(p^+)\rangle,
%---------------------------
%---------------------------
\eeq
%---------------------------

With resort to the completeness relation $\sum |X\rangle \langle X| = 1$, one finds that the summation of the out state
over soft colored spectators in \eqref{eq:csdef} simply reduces to
%---------------------------
\begin{align}
%---------------------------
&\notag \sum_{X}\vert H(P^+)\!+\!X(p^+)\rangle \,\langle X(p^+)\!+\!H(P^+)\vert
\\
&=2P^{+}m_n^{\dagger}\left(P^{+}\right)m_n \left(P^{+}\right).
%---------------------------
\label{eq:sumX}
%---------------------------
\end{align}
%---------------------------
Here the principal quantum number $n$ needs not to be summed.

Substituting \eqref{eq:sumX} into the Collins-Soper definition of FF in \eqref{eq:csdef}, we encounter the vacuum matrix element
$\langle 0 \vert b \, m_n^{\dagger}\left(P^{+}\right) m_n \left(P^{+}\right)\, b^\dagger \vert 0 \rangle $, which turns out to simply vanish.
Therefore the quark fragmentation function vanishes at order-$N_c^0$.

In fact, the out state $|H+X\rangle$ in \eqref{eq:csdef} should be the eigenstate of the full LF Hamiltonian in \eqref{eq:HLF_full},
instead of the leading-color piece $\HLFz$. To get a nonvanishing result, we need incorporating the quantum mechanical correction into
the state $|H+X\rangle$, treating $\rm{V}$ in eq.~\eqref{eq:V} as a first-order correction:
%---------------------------
\begin{align}
%---------------------------
&\vert H(P^+)+X(p^+)\rangle'=
\notag\vert H(P^+)+X(p^+)\rangle
%---------------------------
\\
%---------------------------
&+\frac{1}{P_{H\!+\!X}^{-}-\HLFz+i\epsilon} {\rm V} \vert H(P^+)+X(p^+)\rangle+\cdots,
%---------------------------
\label{eq:L-S}
%---------------------------
\end{align}
%---------------------------
with $\left|H+X\right\rangle'$ denoting the eigenstate of the full LF Hamiltonian.

Note the second piece in \eqref{eq:V} is suppressed by a factor of $1/\sqrt{N_c}$ with the leading Fock state. This corresponds to a process where a gluon is emitted from
the initial quark and the {virtual gluon} then splits into quark-antiquark pair, $q\to qg\to q q \bar{q}$.
One quark will combine with the antiquark to form the identified meson state $\vert H \rangle$,  while the orphan quark transition
into the spectator state $\vert X\rangle$.

Applying \eqref{eq:L-S} for both ket and bra of the asymptotic out states in \eqref{eq:csdef},  we find that the
the leading non-vanishing contribution to the quark FF is of ${\cal O}(1/N_c)$ reads
%---------------------------
\begin{align}
%---------------------------
\notag D_{q\to H}(z)=& \frac{z^{d-3}}{4\pi}\int d\xi^- e^{iz^{-1}P^+\xi^-}\frac{1}{N_c}\mathrm{Tr_{color}Tr_{Dirac}}
%---------------------------
\\
%---------------------------
\notag &\times\left\langle 0\right|\psi_R(0)\frac{1}{P_{H+X}^- \!- \!\HLFz \!+\!i\epsilon} {\rm V} (2P^+)m_n^\dagger(P^+)
%---------------------------
\\
%---------------------------
&\times m_n(P^+) {\rm V} \frac{1}{P_{H+X}^-\! -\! \HLFz\!-\!i\epsilon}\overline{\psi}_R(\xi^-)\left|0\right\rangle,
%---------------------------
\label{eq:FF_recast1}
%---------------------------
\end{align}
%---------------------------
where we have employed \eqref{eq:sumX} to eliminate the sum over the soft spectator states $\vert X \rangle$.

With the aid of the commutation relation \eqref{commutator:HLF:b+}, we make the following simplification in \eqref{eq:FF_recast1}:
%---------------------------
\bseq
%---------------------------
\begin{align}
%---------------------------
 &\frac{1}{P_{H+X}^--\HLFz-i\epsilon}\psi_{R}^{i\dagger}\left(\xi^{-}\right)\vert0\rangle
%---------------------------
= \int_{0}^{\infty}\frac{dk^{+}}{2\pi}e^{ik^{+}\xi^{-}} \frac{1}{P_{H+X}^--\left(\frac{m^2 - 2\lambda}{2k^+} + \frac{\lambda}{\rho}\right)-i\epsilon}
b^{i\dagger}({k^+}) \vert0\rangle,
%---------------------------
\\
%---------------------------
 &\langle0\vert\psi_{R}^i\left(0\right)\frac{1}{P_{H+X}^--\HLFz+i\epsilon} = \int_{0}^{\infty}\frac{dq^{+}}{2\pi}\langle0\vert  b^i({q^+})
\frac{1}{P_{H+X}^--\left(\frac{m^2 - 2\lambda}{2q^+} + \frac{\lambda}{\rho}\right)+i\epsilon},
%---------------------------
\end{align}
%---------------------------
\label{eq:quark_on_vac}
%---------------------------
\eseq
%---------------------------
where $i$ denotes the color index. It is worth noting that, hearteningly,
the dangerous $\lambda/\rho$ term in the energy denominator in \eqref{eq:quark_on_vac} cancels between the
dressed parent quark light-front energy and $P^-_{H+X}$ (recall \eqref{eq:PHX}).

Plugging \eqref{eq:PHX} and \eqref{eq:quark_on_vac} into \eqref{eq:FF_recast1}, we arrive at
%---------------------------
\begin{align}
%---------------------------
    \notag D(z)=&\frac{z^{d-3}}{4\pi}\int d\xi^- e^{iz^{-1}P^+\xi^-}\frac{1}{N_c}\mathrm{Tr_{color}}\mathrm{Tr_{Dirac}}
%---------------------------
\\
%---------------------------
    \notag&\times\int_0^{\infty}\frac{dq^+}{2\pi}\int_0^{\infty}\frac{dk^+}{2\pi}e^{-ik^+\xi^-}8P^+\\&
   \notag \times\frac{1}{\left[(m^2-2\lambda)\left(\frac{1}{q^+}+\frac{1}{P^+-q^+}\right)-\frac{M_H^2}{P^+}+i\epsilon\right]}\\
   \notag &\times\frac{1}{\left[(m^2-2\lambda)\left(\frac{1}{k^+}+\frac{1}{P^+-k^+}\right)-\frac{M_H^2}{P^+}-i\epsilon\right]}\\
    & \times\left\langle 0\right|b^i\left(q^+\right)\,{\rm V}\, m_n^\dagger(P^+)m_n(P^+)\, {\rm V} \,b^{i\dagger}\left(k^+\right)\left|0\right\rangle,
%---------------------------
\label{eq:FF_recast2}
%---------------------------
\end{align}
%---------------------------

The vacuum matrix element in \eqref{eq:FF_recast2} can be worked out by plugging the explicit definition of the
color-suppressed operator ${\rm V}$ in \eqref{eq:V}, and making use of the following
commutation relations among $b$, $B$, $D$, $M$, $m$
%---------------------------
\bseq
%---------------------------
\begin{align}
  &\left[b^i\left(p^+\right), B\left(q^+, k^+\right)\right] =
    2\pi\delta\left(p^+-q^+\right) b^i\left(k^+\right), \\
  &\notag \left[M\left(p^+, q^+\right), m_n^\dagger\left(P^+\right)\right]\\
   &=
    \sqrt{\frac{2\pi}{P^+}} \phi_n\left(\frac{q^+}{P^+}\right)2\pi\delta
    \left(P^+-p^+-q^+\right)+\mathcal{O}(N_c^{-1}).
\end{align}
%---------------------------
\eseq
%---------------------------
Integrating over $\xi^-$ and $q^+$ and $k^+$ in \eqref{eq:FF_recast2},
we are able to deduce the functional form of the quark-to-meson FF:
%---------------------------
\beq
%---------------------------
D_{q\rightarrow H}(z) = \frac{4\lambda^2}{N_c}\frac{(1-z)^2}{z\left[\left(z^2(m^2-2\lambda)+(1-z)M_H^2\right)^2+\epsilon^2\right]}\left[\int_0^1 dy\,\frac{\phi_n(y)}{\left(y-1/z\right)^2}\right]^2,
%---------------------------
\label{eq:FF_cs}
%---------------------------
\eeq
%---------------------------
which does scale as ${\cal O}(1/N_c)$.
Note only a single 't Hooft wave function $\phi_n(y)$ associated with the daughter meson is invoked in \eqref{eq:FF_cs},
which looks much simpler than the expression of the FF given by Ellis~\cite{Ellis:1977mw}.
Not surprisingly, the functional form the fragmentation function is considerably more involved than the quark PDF of a meson
in 't Hooft model, which is simply $|\phi_n(x)|^2$.

Equation~\eqref{eq:FF_cs} is the main result of this work.
For a consistency check, we also provide an alternative derivation of this result from diagrammatic approach in Appendix.

Inspecting \eqref{eq:FF_cs}, one observes that, irrespective of the quark mass,
the quark FF $D_{q\rightarrow H}(z)$ always vanishes at end points $z=0,\:1$.

In the heavy quark case ($m>\sqrt{2\lambda}$), the terms inside the parenthesis in the denominator in \eqref{eq:FF_cs} are all positive definite
in entire regime of $z$, therefore it is safe to drop the infinitesimal $\epsilon$ factor.

Unfortunately, some pathetic behavior emerges in the light quark case ($m<\sqrt{2\lambda}$). Because the denominator in \eqref{eq:FF_cs} may vanish
in some specific values of $z$, the light quark FF simply blows up. This symptom might originate
from the fact that the light quark becomes tachyonic (the renormalized quark mass squared becomes negative) in the 't Hooft model, whose light-front energy becomes negative~\cite{Coleman:1985rnk,Manohar:1998xv}.
This is in conflict with the probability interpretation of the quark FF, which characterizes
the probability density of a single quark fragmenting into an identified hadron accompanied with soft spectators.
We hope that the future study can shed some light on how to ameliorate the divergent behavior of the light quark FF.

In the next section, we will use an alternative first-principle method (NRQCD approach) to test the correctness
of \eqref{eq:FF_cs}. Therefore, in the rest of the paper, we will solely focus on the heavy quark sector, 
where \eqref{eq:FF_cs} gives an unambiguous and well-behaved description
of the heavy quark fragmentation function. 

\section{heavy quark fragmentation function from NRQCD factorization}
\label{sec:nrqcd}

One may naturally wonder how to test the correctness of \eqref{eq:FF_cs}? Needless to say, one cannot resort to experimental input.
It would be highly desirable if there exists some alternative nonperturbative method to compute the quark fragmentation function,
which can be used to test against \eqref{eq:FF_cs}.

Fortunately, in the realistic world, there exists a robust field-theoretical approach to describe the fragmentation function for heavy quarkonium, {\it i.e.},
the NRQCD factorization approach. Unlike the parton fragmentation into light hadrons,
the fragmentation functions for heavy quarkonium needs not to be a genuinely nonperturbative object.
Owing to $m\gg \Lambda_{\rm QCD}$, one can invoke the asymptotic freedom to further factorize the heavy quark to quarkonium fragmentation function into
the sum of products of short-distance coefficients (SDCs) and long-distance NRQCD matrix elements (LDMEs)~\cite{Bodwin:1994jh}.
At the lowest order in velocity expansion, the heavy quark fragmentation function can be expressed as~\footnote{In this section, we use the symbol $Q$ to denote
the heavy quark.}
%---------------------------
\begin{align}
%---------------------------
D_{Q\rightarrow H}(z)= d^{(0)}(z)\left\langle 0\vert \mathcal{O}_1^H\right \vert 0 \rangle+\mathcal{O}\left(v^2\right),
%---------------------------
\label{eq:NRQCD_fact_1}
%---------------------------
\end{align}
%---------------------------
where $d^{(0)}$ denotes the leading-order SDC. The leading-order NRQCD production operator $\mathcal{O}_1^{H}$ is defined by
%---------------------------
\begin{align}
%---------------------------
\mathcal{O}_1^H = &\sum_X \chi^\dagger\psi\left|H+X\right\rangle\left\langle X+H\right| \psi^\dagger \chi,\label{eq:NRQCD_O}
%---------------------------
\end{align}
%---------------------------
where $\psi$/$\chi$ are Pauli spinor fields that annihilate/create heavy quark/antiquark.
$X$ represents the additional soft spectator hadrons.
Since $\chi^\dagger\psi$ is a color-singlet operator, hence $X$ must be color-singlet state.

Since 't Hooft model also exhibits ``naive" asymptotic freedom, we will take the validity of NRQCD factorization for granted
when applied to 1+1 dimensional QCD.
Concretely speaking, the theme of this section is to recalculate the heavy quark fragmentation function using \eqref{eq:NRQCD_fact_1} in $\QCDtwo$~\footnote{Note that in
NRQCD factorization approach, one does not need assuming the $N_c\to \infty$ limit.}.
{For simplicity, we will concentrate on the heavy quark fragmentation into the ground state quarkonium with $n=0$.
}

\subsection{Determination of the SDC through perturbative matching}

We use the standard perturbative matching approach to determine $d^{(0)}$.
Since the SDC is insensitive to long-distance physics, one replaces the physical quarkonium $|H\rangle$
by a fictitious quarkonium, {\it i.e.}, the free quark-antiquark pair $|q\bar{q}\rangle$, where each of the
constitutes share equal momentum $P/2$. {One then has $P^2= 4 E^2\approx 4 m^2$.}
For a heavy quark fragmenting into a fictitious quarkonium,
the NRQCD factorization formula \eqref{eq:NRQCD_fact_1} can be recast into
%---------------------------
\begin{align}
%---------------------------
D_{Q\rightarrow Q\overline Q}(z)=d^{(0)}(z)\left\langle \mathcal{O}^{Q\overline Q}_1\right\rangle+\mathcal{O}\left(v^2\right).
%---------------------------
\label{eq:D_qq_fact1}
%---------------------------
\end{align}
%---------------------------
By calculating the left-hand side using perturbative QCD, and calculating the right-hand side using perturbative NRQCD, one can readily solve
$d^{(0)}(z)$.

\begin{figure}[htbp]
\begin{center}
\includegraphics[clip,width=0.4\textwidth]{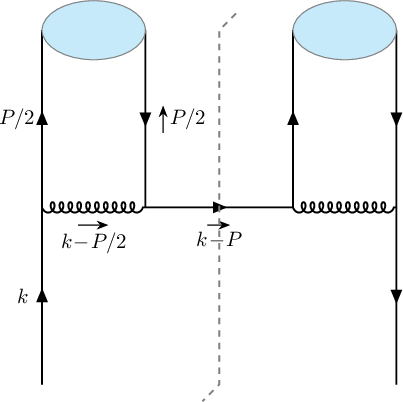}
\caption{Tree-level diagram for the quark-level fragmentation function for $Q\rightarrow Q\bar{Q}$ in light-cone gauge.
The shaded bubble represents the quarkonium and the vertical dashed line represents imposing a cut.}
\label{fig:q_to_qqbr_FF}
\end{center}
\end{figure}

Again we start from the Collins-Soper definition \eqref{eq:csdef} to compute $D_{Q\rightarrow Q\overline Q}(z)$.
It is convenient to work in light-cone gauge $A^+\!=\!0$, so that the gauge link $\mathcal{W}$ in \eqref{eq:csdef} can be discarded.
The tree-level Feynman diagram for $Q\rightarrow Q\overline{Q}$ is shown in Fig.~\ref{fig:q_to_qqbr_FF}.

Following the Feynman rules of the cut diagram in Fig.~\ref{fig:q_to_qqbr_FF}, we have
%---------------------------
\begin{align}
%---------------------------
\notag D_{Q\rightarrow Q\bar{Q}}(z)=&\frac{g_s^4 C_F^2}{\pi N_c}\frac{z^{-1}E}{2(k^+-P^+)}\\
&\left. \times\mathrm{Tr_{Dirac}}\left[\gamma^+ \mathcal{A}^\dagger\left(k\!\!\!/-P\!\!\!\!/+m\right)\mathcal{A} \right] \right|_{P^+=zk^+},
%---------------------------
\label{eq:D_q_to_qqbr}
%---------------------------
\end{align}
%---------------------------
where the amplitude $\mathcal{A}$ reads
%---------------------------
%---------------------------
\begin{align}
%---------------------------
\mathcal{A} = &\frac{
  \gamma^\mu v\left(\frac{P}{2}\right)\bar{u}\left(\frac{P}{2}\right)
  \gamma^\nu\left(k\!\!\!/+m\right)
  }{
    (k^2-m^2)\left(k-\frac{P}{2}\right)^2
  }\left[
    \notag g_{\mu\nu}-\frac{\big(k_\mu-\frac{P_\mu}{2}\big)n_\nu}{k^+-\frac{P^+}{2}}\right.
%---------------------------
\\
%---------------------------
&\left.-\frac{\big(k_\nu-\frac{P_\nu}{2}\big)n_\mu}{k^+-\frac{P^+}{2}}
\right].
%---------------------------
\end{align}
%---------------------------
%---------------------------
The auxiliary 2-vector $n_\mu=(1,-1)/\sqrt{2}$ denotes the null vector along the ``-'' direction,
which satisfies $n^2=0$ and $k\cdot n =k^+$.

The physical results should not depend on the convention adopted for Dirac $\gamma$ matrices. However,
to deal with nonrelativistic fermion-antifermion system, it is advantageous in this section to temporarily switch to the Dirac-Pauli basis:
%---------------------------
\begin{equation}
    \gamma^0=\sigma_3,\quad \gamma^1= i\sigma_2,\quad \gamma_5\equiv \gamma^0\gamma^1= \sigma_1,
\end{equation}                       
%---------------------------
In the Dirac-Pauli basis, the Dirac spinor wave functions become
%---------------------------
\beq
u\left(p\right) = \frac{1}{\sqrt{2E}}\left( \begin{array}{cc}
  \sqrt{E + m}\vphantom{\frac{1}{\int}}  \\
  \sqrt{E - m}\vphantom{\frac{\int}{1}}
  \end{array}   \right),\;\;
v\left(p\right)=\frac{1}{\sqrt{2E}}\left( \begin{array}{cc}
  \!\!-\sqrt{E - m}\vphantom{\frac{1}{\int}}
%---------------------------
\\
%---------------------------
\sqrt{E + m}\vphantom{\frac{\int}{1}}
\end{array}   \right).
%---------------------------
\eeq
%---------------------------
Note that the Dirac spinors are chosen to be normalized non-relativistically, {\it i.e.},
$\bar{u}\left(p\right)u\left(p\right)=-\bar{v}\left(p\right)v\left(p\right) = {m/E}$.

After some straightforward algebra, we obtain the resulting perturbative quark-level fragmentation function:
%---------------------------
\begin{align}
%---------------------------
D_{Q\rightarrow Q\bar{Q}}(z) = \frac{64\pi\lambda^2\left(N_c^2-1\right)^2}{m_Q^5N_c^5}\frac{z^3(1-z)^2}{(2-z)^8}+\cdots.
%---------------------------
\label{eq:Dqq_res}
%---------------------------
\end{align}
%---------------------------

We then compute the production matrix element $\left\langle \mathcal{O}_1^{Q\bar{Q}} \right\rangle$ in \eqref{eq:D_qq_fact1} in perturbative NRQCD.
At lowest order in velocity, it is legitimate to drop the soft spectator $X$ by invoking vacuum saturation approximation:
%---------------------------
\begin{align}
%---------------------------
\left\langle \mathcal{O}_1^{Q\bar{Q}} \right\rangle = \left\langle 0\vert \chi^\dagger\psi \vert  Q\bar{Q} \right\rangle \left\langle  \bar{Q}Q \vert \psi^\dagger \chi\vert  0
\right\rangle,
%---------------------------
\end{align}
%---------------------------

The NRQCD fields can be Fourier-expanded as
%---------------------------
\begin{align}
%---------------------------
  \psi^i\left(x\right) = \int\frac{dk}{2\pi}\, e^{i k x} \tilde{b}_{k}^{i},\;\;
  \chi^i\left(x\right) = \int\frac{dk}{2\pi}\, e^{-i k x} \tilde{d}^{i\dagger}_{k}.
%---------------------------
\label{NRQCD:field:Fourier:expand}
%---------------------------
\end{align}
%---------------------------

We are working in the rest frame of the fictitious quarkonium state. At lowest order in velocity, both heavy quark and heavy antiquark inside
this fictitious quarkonium are at rest:
%---------------------------
\begin{align}
%---------------------------
\left\vert Q\overline{Q} \right\rangle = \frac{1}{\sqrt{N_c}} \tilde{b}^{i\dagger}_{0} \tilde{d}^{i\dagger}_{0}\vert 0\rangle.
%---------------------------
\label{fictitious:quarkonium:at:rest}
\end{align}
%---------------------------

Combining \eqref{NRQCD:field:Fourier:expand} and \eqref{fictitious:quarkonium:at:rest}, we immediately obtain
$\left\langle 0 \left\vert \chi^\dagger\psi \right\vert Q\bar{Q} \right\rangle = \sqrt{N_c}$, so that
the desired NRQCD production matrix element becomes~\footnote{Note the counterpart of this NRQCD matrix element
in $\QCDfour$ has an extra factor of 2, stemming from the sum over spin degrees of freedom, which is absent in $\QCDtwo$.}
%---------------------------
\begin{align}
%---------------------------
    \left\langle \mathcal{O}_1^{Q\bar{Q}} \right\rangle = N_c.
%---------------------------
\label{eq:qq_NRQCDME}
%---------------------------
\end{align}
%---------------------------

Substituting \eqref{eq:Dqq_res} and \eqref{eq:qq_NRQCDME} into \eqref{eq:D_qq_fact1}, one can readily solve the
desired SDC:
%---------------------------
\begin{align}
%---------------------------
d^{(0)}(z) =& \frac{64\pi\lambda^2}{m^5N_c^2}\frac{z^3(1-z)^2}{(2-z)^8}+{\cal O}(1/N_c^4).
%---------------------------
\end{align}
%---------------------------
In order to compare with \eqref{eq:FF_cs}, we have retained the leading color piece in the SDC.

Therefore, the quark-to-quarkonium FF in NRQCD factorization reads
%---------------------------
\begin{align}
%---------------------------
D_{Q\rightarrow H}(z)=  \frac{64\pi\lambda^2}{m^5N_c^2}\frac{z^3(1-z)^2}{(2-z)^8}\left\langle \mathcal{O}_1^H\right\rangle+
\mathcal{O}\left(v^2\right).
%---------------------------
\label{eq:NRQCD_FF}
%---------------------------
\end{align}
%---------------------------
As we will show in next subsection, the nonperturbative NRQCD production matrix element $\left\langle \mathcal{O}_1^H\right\rangle$
scales as ${\cal O}(N_c)$, so the heavy quark fragmentation function exhibits an overall $1/N_c$ scaling, compatible with
what is found in the {\it ab initio} derivation, \eqref{eq:FF_cs}.

\subsection{Determination of NRQCD LDME}
\label{sec:LDME}

To make concrete prediction from \eqref{eq:NRQCD_FF}, we need know the actual value of the leading-order NRQCD production matrix element.
Obviously one needs some nonperturbative means to determine its value.
First we note that, at the lowest order in $v$, it is legitimate to invoke the vacuum saturation approximation to
reexpress the vacuum expectation value of \eqref{eq:NRQCD_O} as
%---------------------------
\beq
 \langle\mathcal{O}_1^H\rangle \approx  \left|\left\langle 0\vert \chi^\dagger\psi \vert  H\right\rangle\right|^2.\label{eq:NRQCD_O_vac}
%---------------------------
\eeq
%---------------------------

In the following, we use two different approaches to estimate the value of the vacuum-to-quarkonium matrix element
$\langle 0\vert \chi^\dagger\psi \vert  H\rangle$.

\subsubsection{Estimation from the decay constant}

First define the decay constant of the ground-state ($n=0$) quarkonium through
%---------------------------
\begin{align}
%---------------------------
\left\langle 0 \left| \overline{\Psi}\gamma^\mu\gamma^5 \Psi\right|H(P)\right\rangle \equiv {1\over \sqrt{2P^0}} f_H P^\mu,
\label{eq:fH_def}
%---------------------------
\end{align}
%---------------------------
 where $\Psi$ denotes the Dirac field for heavy quark $Q$.

The decay constant $f_H$ is related to the first moment of the 't Hooft wave function through~\cite{Callan:1975ps}
%---------------------------
\begin{align}
%---------------------------
f_H = \sqrt{\frac{N_c}{\pi}}\int_0^1 dx \phi_0(x),
\label{eq:fH_LCWF}
%---------------------------
\end{align}
%---------------------------
which scales as $\sqrt{N_c}$.

Since the decay constant is a Lorentz scalar, we can compute it in any reference frame, {\it e.g.}, the rest frame of the quarkonium.
One can match the QCD axial vector current $\overline{\Psi}\gamma^\mu\gamma^5 \Psi$ onto NRQCD bilinear $\chi^\dagger \psi$ by integrating out
relativistic fluctuation. A shortcut to carry out the tree-level matching is through the Foldy-Wouthuysen-Tani (FWT) transformation:
%---------------------------
\begin{align}
%---------------------------
\Psi = \exp\left(-\frac{i}{2m_Q}\gamma^1 D_1 \right)\left(
  \psi,  \chi \right)^\mathrm{T},
%---------------------------
\end{align}
%---------------------------
where $D_1$ signifies the spatial component of the covariant derivative.

Performing FWT transformation to the QCD axial vector current in \eqref{eq:fH_def} (taking $\mu=0$ component),
we find
%---------------------------
\beq
%---------------------------
f_H = \sqrt{\frac{2}{M_H}}\left(1+{\cal O}\left(g_s^2/ m^2\right) \right)\langle 0\vert\chi^\dagger\psi\vert H(P=0)\rangle + \mathcal{O}(v^2),
%---------------------------
\label{eq:fH_NRQCD}
%---------------------------
\eeq
%---------------------------
at the lowest order in relative velocity.
We caution that we have not considered the radiative corrections to the SDC,
which may induce the correction of the relative order $g_s^2/m_Q^2$.

Piecing together \eqref{eq:NRQCD_O_vac}, \eqref{eq:fH_LCWF} and \eqref{eq:fH_NRQCD}, we
are able to express the desired NRQCD LDME in terms of the first momentum of the 't Hooft wave function:
%---------------------------
\beq
%---------------------------
\langle\mathcal{O}_1^H\rangle =  \frac{M_H}{2}f_H^2 =  \frac{N_c }{2\pi} M_H \left[\int_0^1 dx\, \phi_0(x)\right]^2.
%---------------------------
\label{eq:O_decay}
%---------------------------
\eeq
%---------------------------

\subsubsection{Estimation from Schr\"{o}dinger wave function at the origin}

In realistic ${\rm QCD}_4$, the vacuum-to-quarkonium NRQCD matrix element is often approximated by the wave function at the origin in phenomenological
potential model such as Cornell model.
For instance, the color-singlet NRQCD production matrix element for $\eta_c({}^1S_0)$ can be approximated by
$\langle \mathcal{O}_1^{\eta_c} \rangle\approx \frac{2}{4\pi} N_c | R(0)|^2$, where the $R(0)$ denotes the radial wave function at origin for $\eta_c$.
The factor $1/(4\pi)$ is the spherical harmonic $Y_{00}(\hat{\boldsymbol{r}})$ and the factor $2$ originates from the sum over spin degree of freedom~\cite{Bodwin:1994jh}.

In 1+1 dimensional QCD, the instantaneous gluon exchange naturally leads to a linear Coulomb potential. Analogous to the ${\rm QCD}_4$,
we may also approximate the NRQCD matrix element of the ground-state quarkonium
by the wave function at the origin:
%---------------------------
\begin{align}
%---------------------------
\left\langle \mathcal{O}^{H}_1\right\rangle \approx {N_c}\left| \psi_H(0) \right|^2.
%---------------------------
\label{eq:O_to_WF}
%---------------------------
\end{align}
%---------------------------
The difference of this expression from its four-dimensional counterpart originates from the fact there is no concept of angular momentum (orbital or spin)
in 1+1 dimensional spacetime.

In contrast to the light-front quantization, there is also possible to use the equal-time quantization
(together with imposing the axial gauge) to solve the 't Hooft model~\cite{Bars:1977ud}.
The resulting field-theoretical bound-state equations, the so-called Bars-Green equations, are considerably more involved than the 't Hooft equation,
and are suitable for describing bound state carrying any {\it finite} momentum.
In the $m_Q\to \infty$ limit, the heavy quarkonium becomes a nonrelativistic system. One can rigorously proves that,
in the heavy quark limit, the Bars-Green function in the quarkonium rest frame reduces to the following form~\cite{Jia:2018mqi}
%---------------------------
\beq
%---------------------------
\frac{p^2}{2\mu}\varphi^{n}_+(p)-\lambda\dashint_{-\infty}^{\infty}\frac{dk}{(p-k)^2}\varphi^{n}_+(k)=\mathcal{E}_n\varphi^{n}_+(p),
%---------------------------
\eeq
%---------------------------
where $\mu=m_Q/2$ signifies the reduced mass. $\mathcal{E}_n$ denotes the binding energy $\mathcal{E}_n=M_H-2m_Q$, while
$n$ is the principal quantum number of quarkonium. After Fourier transformation, this equation is nothing but the
familiar Schr\"odinger equation with a linear potential in position space:
%---------------------------
\begin{align}
%---------------------------
-\frac{1}{2\mu}\frac{d^2}{dr^2} \psi_H(r) +\lambda\pi |r| \psi_H(r) = \mathcal{E}_n\psi_H(r).~\label{eq:schrodinger}
%---------------------------
\end{align}
%---------------------------
We stress again, in contrast to the 3+1 dimension, this Schr\"odinger equation arises from a rigorous field-theoretical
bound-state equation rather from a phenomenological model. We refer interested readers to \cite{Jia:2018mqi} for detailed discussions.

This Schr\"odinger equation can be readily solved. The ground-state eigenenergy can be determined by the condition
$\mathrm{Ai}'\!\left(-\frac{\mathcal{E}_{0}(2 \mu)^{1 / 3}}{(\pi \lambda)^{2 / 3}}\right)=0$, with
with $\mathrm{Ai}$ denoting the Airy function.
The  Schr\"odinger wave function of the ground-state quarkoniunm is~\cite{Jia:2018mqi}
%---------------------------
\beq
%---------------------------
\psi_{H}(r) =  \mathcal{N}_{0} \mathrm{Ai}\left(\frac{\mu^{1 / 3}\left(-2 \pi \lambda |r|-2 \mathcal{E}_{0}\right)}{(2 \pi \lambda)^{2 / 3}}\right),
%---------------------------
\label{wv:func:Schroedinger}
\eeq
%---------------------------
where the normalization constant $\mathcal{N}_0$ is determined by the requirement $\int dr\, \psi^2_H(r)=1$.
Clearly the ground-state quarkonium carry even parity, {\it i.e.}, $\psi(r)=\psi(-r)$, therefore its wave function
at the origin does not vanish.

Substituting \eqref{wv:func:Schroedinger} into \eqref{eq:O_to_WF}, we obtain another estimation of the NRQCD production matrix element:
%---------------------------
\begin{align}
%---------------------------
\left\langle \mathcal{O}^{H}_1\right\rangle \approx {N_c}\,\mathcal{N}_0^2
\left[\mathrm{Ai}\left(\frac{-2\mathcal{E}_0\mu^{1/3}}{\left(2\pi\lambda\right)^{2/3}}\right)\right]^2.
\label{eq:LDME}
%---------------------------
\end{align}
%---------------------------

\section{Compatibility test between NRQCD factorization and rigorous results in heavy quark limit}
\label{comparison:two:approaches}

We now try to examine whether the {\it ab initio} result for the heavy quark fragmentation function, \eqref{eq:FF_cs}, is compatible with the
NRQCD prediction, \eqref{eq:NRQCD_FF}. To make such a comparison meaningful, we need to take the heavy quark limit $m\gg \sqrt{2\lambda}$.
In such a limit, $M_H\approx2m$ is a very decent approximation. Intuitively speaking,
as the quark mass increases, the bound state becomes more and more non-relativistic,
so that the heavy quark and heavy anti-quark almost equally partition the heavy meson's momentum for a fast-moving quarkonium.
In the so-called impulse approximation, the LCWF of the ground-state quarkonium $H$ can be approximated by
%---------------------------
\beq
%---------------------------
  \phi_0(x)\approx \sqrt{\frac{\pi}{N_c}}f_H\delta\left(x-\frac{1}{2}\right).
%---------------------------
\label{impulse:appxo}
%---------------------------
\eeq
%---------------------------
The shrinking tendency of the profile of the LCWF with increasing quark mass has been numerically exhibited~\cite{Jia:2017uul}.
The normalization factor is chosen such that \eqref{eq:fH_LCWF} is fulfilled.

Substituting the idealized LCWF of the ground-state quarkonium in \eqref{impulse:appxo} into our {\it ab initio} formula, \eqref{eq:FF_cs}, we obtain
%---------------------------
\begin{align}
%---------------------------
D_{Q\rightarrow H}(z) \approx &\frac{{4}\lambda^2}{N_c}\frac{(1-z)^2}{z\left[z^2 m^2+(1-z)M_H^2\right]^2}
%---------------------------
\left[\frac{1}{\left(1/2-1/z\right)^2}\right]^2
\frac{\pi}{N_c}f_H^2.
%---------------------------
\label{eq:impluse}
%---------------------------
\end{align}
%---------------------------

Hearteningly, with the aid of the relation \eqref{eq:O_decay}, we find \eqref{eq:impluse} is in exact agreement with the lowest-order NRQCD
prediction \eqref{eq:NRQCD_FF}.

In our opinion, this agreement is by no means accidental, since both methods have completely different origins. We believe our field-theoretical expression for
heavy quark fragmentation function in 't Hooft model, \eqref{eq:FF_cs}, is robust and correct.

\section{Numerical results}\label{sec:numerical}

\begingroup
\setlength{\tabcolsep}{10pt} % Default value: 6pt
\renewcommand{\arraystretch}{1.5}
\begin{table}[htbp]
\centering
\begin{tabular}{cccc}
\hline\noalign{\smallskip}
& $c$    & $b$     & $b'$      \\
\noalign{\smallskip}\hline\noalign{\smallskip}
$m/\sqrt{2\lambda}$      & 4.19 & 13.66 & 27.32  \\
$M_H/\sqrt{2\lambda}$ & 9.03 & 27.82 & 55.06 \\
$\mathcal{E}_0/\sqrt{2\lambda}$ & 0.854 & 0.576 & 0.457 \\
$\mathcal{N}_0^2/\sqrt{2\lambda}$ & 3.21 & 4.75 & 5.99 \\
$\left\langle \mathcal{O}_1^H\right\rangle/(N_c\!\sqrt{2\lambda}),\;\text{eq.}~\eqref{eq:LDME}$ & 0.920 & 1.36 & 1.72 \\
$\left\langle \mathcal{O}_1^H\right\rangle/(N_c\!\sqrt{2\lambda}),\;\text{eq.}~\eqref{eq:O_decay}$ & 0.817 & 1.32 & 1.70 \\
\noalign{\smallskip}\hline
\end{tabular}
\caption{The numerical values of the heavy quark masses ($m$), lowest-lying quarkonia masses ($M_H$),
binding energies ($\mathcal{E}_0$), the normalization constant of the lowest-lying quarkonum's wave function ($\mathcal{N}_0$),
and the corresponding NRQCD LDMEs estimated by two methods.}\label{tab:mass}
\end{table}
\endgroup

In this section, we present our numerical results of the fragmentation functions for three characteristic heavy quarkonia:
$c\bar{c}$, $b\bar{b}$ and a fictitious heavier quarkonium $b'\bar{b'}$ with $m_b'=2m_b$.  The value of 't Hooft coupling constant $\lambda$ is chosen as
$\sqrt{2\lambda}=340\:\mathrm{MeV}$, compatible with the empirical value of the string tension in the $\QCDfour$~\cite{Burkardt:2000uu}.
The quark masses are tuned so that the masses of the ground state charmonium and bottomonium coincide with those of $J/\psi$ and $\Upsilon$ in realistic world.
For more details about quark mass setting, we refer the interested readers to Ref.~\cite{Jia:2017uul}.
The masses of three heavy quarks, $c$, $b$, $b^\prime$, and the respective ground-state quarkonia,
together with the respective binding energies, are tabulated in Table~\ref{tab:mass}.
All the energy and mass are given in units of $\sqrt{2\lambda}$.

\begin{figure}[htbp]
\begin{center}
\resizebox{0.45\textwidth}{!}{\includegraphics{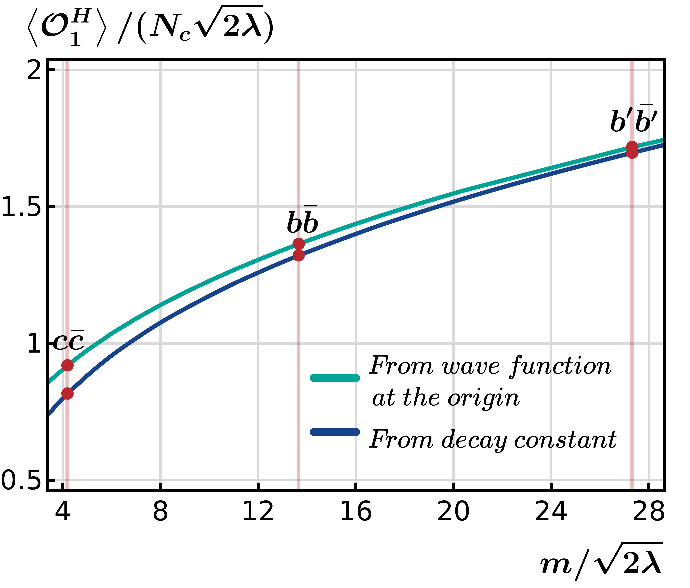}}
\caption{The NRQCD production matrix element $\left\langle \mathcal{O}^{H}_1\right\rangle$ as function of heavy
quark mass. Three benchmark quarkonia $c\bar{c}$, $b\bar{b}$ and $b\bar{b}'$ are marked with heavy dots.}
\label{fig:nrqcdme}
\end{center}
\end{figure}

In Fig.~\ref{fig:nrqcdme}, we plot the NRQCD production matrix element as a function of quark mass, obtained from two different ways, {\it i.e.},
the estimation from the wave function at the origin
and estimation from the decay constant. The discrepancy for the moderate quark mass might be attributed to the neglected QCD
radiative correction of ${\cal O}(g_s^2/m^2)$, as well as the neglected relativistic correction
upon matching the decay constant $f_H$ onto NRQCD matrix element (see \eqref{eq:fH_NRQCD}). Nevertheless, it is clear in Fig.~\ref{fig:nrqcdme}
these two estimations start to coincide in the heavy quark limit.
We also observe that the values of the NRQCD LDMEs and the normalization constant of the wave function increase with the increasing quark mass.
This tendency simply reflects the fact that the non-relativistic quark-antiquark bound state becomes more spatially concentrated as quark mass increases.

\begin{figure*}[htbp]
\begin{center}
\includegraphics[clip,width=0.95\textwidth]{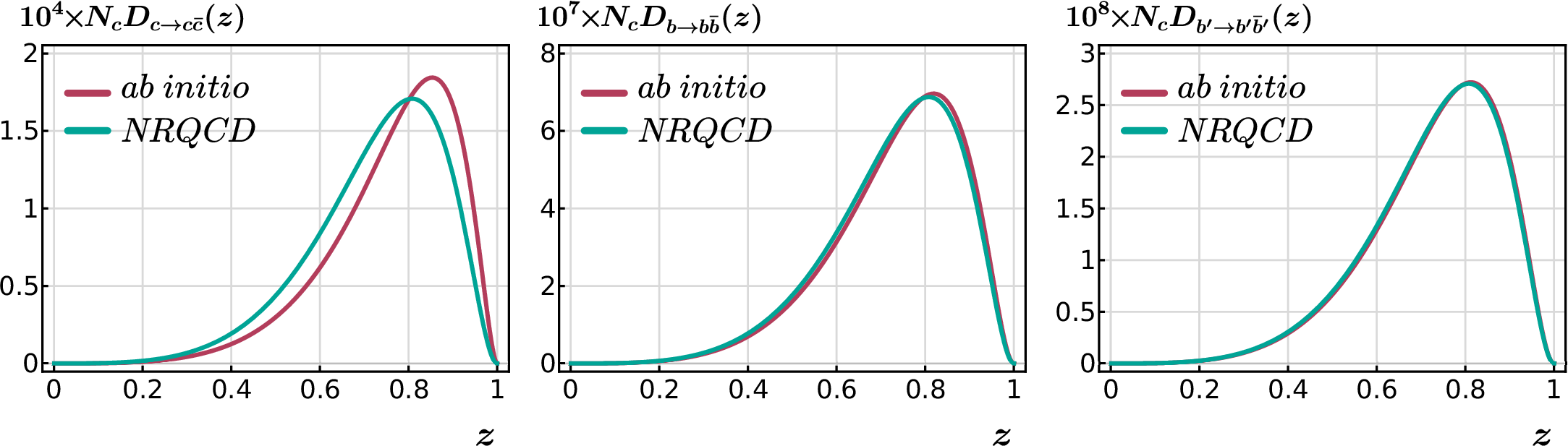}
\caption{Comparison between heavy quark fragmenation functions obtained by the rigorous approach (labeled as {\it ab initio}) and NRQCD factorization approach, for three
benchmark quarkonia.}
\label{fig:FF}
\end{center}
\end{figure*}

Most importantly, in Fig.~\ref{fig:FF} we also show the profiles of heavy quark fragmentation functions for $c\bar{c}$, $b\bar{b}$ and $b^\prime \bar{b}^\prime$,
obtained both from the {\it ab initio} and NRQCD approaches. For the latter, we adopt the value of the LDME estimated form the wave function at the origin.
For charm quark fragmentation, we observe some discrepancy between two approaches, which can be attributed to the fact the charm quark is not very heavy.
Featherless, as can be clearly visualized from Fig.~\ref{fig:FF}, as the quark mass continues to increase,
the convergence between two predictions becomes perfect. This numerical verification supports the
analytical proof of the compatibility between two approaches in the heavy quark limit in Sec.~\ref{comparison:two:approaches}.

\section{Summary and Outlook}\label{sec:sum}

Fragmentation functions are important theoretical ingredients to describe the identified hadron production in high energy collision experiments.
A better understanding of fragmentation function will deepen our knowledge about hadronization mechanism and color confinement.
Unfortunately, in the forseeable future, there seems no reliable first-principle method to compute the fragmentation function in realistic ${\rm QCD}_4$.
In this work, within the framework of the 't Hooft model, a famous toy model of QCD, we attempt to compute the quark-to-meson fragmentation function in a rigorous way,
with the hope that some useful lessons will be gleaned from this study.

Following Collins-Soper definition, we are able to establish the functional form of the quark-to-meson fragmentation function in terms of the identified meson's light-cone wave function,
using both Hamiltonian approach and diagrammatic approach. Its functional form is much simpler than what is given by Ellis about half a century ago.
For comparison, we also employ the NRQCD factorization approach to compute the heavy quark fragmentation into the ground-state quarkonium, at the lowest order in
velocity and strong coupling constant expansion. We show that, both analytically and numerically, as the quark mass increases, the heavy quark fragmentation functions obtained from
the Hamiltonian approach and NRQCD approach converge to each other. Therefore, we believe that both approaches give reliable account of the heavy quark fragmentation in
1+1 dimensional QCD.

We also find some pathetic symptom for the light quark fragmentation in our {\it ab intio} approach,
while the quark-to-meson fragmentation function becomes singular at some value of $z$.
This phenomenon might be intimately linked with the fact that the light quark becomes tachyonic, with negative light-front energy.
Hence the probabilistic interpretation of the quark FF might break down.
It is definitely worth making further efforts to resolve this dilemma, which will deepen our understanding toward
the 't Hooft model and even fragmentation mechanism.

\section*{Acknowledgements}
The work of Y.~J. and Z.-W.~M. is supported in part by the National Natural Science Foundation of China
under Grants No.~11925506,
No.~12070131001 (CRC110 by DFG and NSFC).
The work of X.-N.~X. is supported in part by the National Natural Science Foundation of China under Grant No.~11905296, No.~12275364.

\appendix\label{sec:diagram_approach}
\section*{Appendix: Rederiving quark fragmentation function using diagrammatic approach}

\begin{figure*}\centering
\includegraphics[clip,width=0.6\textwidth]{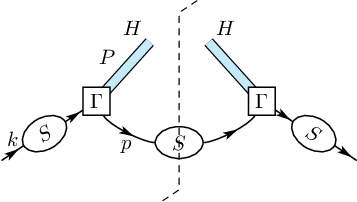}
\caption{The Feynman diagram for quark fragmentation function. }
\label{fig:feyn_diag_FF}
\end{figure*}

In this appendix, we present an alternative approach, namely the Feynman diagram approach to derive the quark FF in 't Hooft model. We show that the quark FFs obtained from 
the diagrammatic approach and the Hamiltonian approach are identical.

The Feynman diagram of a quark fragments into a meson is depicted in Fig.~\ref{fig:feyn_diag_FF}, in which $S$ represents a dressed quark propagator and $\Gamma$ represents the quark-antiquark-meson vertex function. We work in the light-cone gauge. In the framework of LF quantization, the dressed quark propagator $S$ is defined as the time-ordered correlation function of the  ``good'' component of the quark field
\begin{align}
  \langle0\vert\mathrm{T}\psi_R\left(x\right)\psi_R^\dagger\left(y\right)\vert0\rangle &= \int \frac{d^2p}{\left(2\pi\right)^2}\, e^{-ip\cdot\left(x-y\right)}  S_F\left(p\right) \times\mathbf{1}_{\mathrm{color}}
  \end{align}
whose explicit forms is given by~\cite{tHooft:1974pnl}
  \begin{align}
  S_F(p)&= \frac{2ip^+}
    {p^2 -m^2 + 2\lambda - 2\left|p^+\right|\frac{\lambda}{\rho}+i\epsilon},
   \label{eq:dressed_quark_prop}
\end{align}
where $\boldsymbol{1}_{\mathrm{color}}$ denotes a unit operator in color space.

The vertex function $\Gamma$ was introduced in Ref.~\cite{Callan:1975ps,Einhorn:1976ev}, whose form is given by
\begin{align}
 \Gamma \left(xP^+; P^+\right) = &\sqrt{\frac{4\pi}{N_c}}\frac{i\lambda}{\left|P^+\right|}
    \int_0^1 dy\:\Theta\left(\left|x-y\right|-\frac{\rho}{\left|P^+\right|}\right)\frac{\phi_n\left(y\right)}{\left(x-y\right)^2},
\label{eq:Hqqbar_vertex}
\end{align}
where $P^+$ is the light-cone momentum of the meson $H$ and $x$ denotes the
ratio between the light-cone momentum carried by the quark $q$ and meson $H$.

In accordance with Collins-Soper definition of the quark FF, as depicted in Fig.~\ref{fig:feyn_diag_FF}, 
we write down
\begin{align}
\notag  D_{q\to H}\left(z\right) =& \frac{1}{4\pi z}
    \int_0^{\infty}\frac{d^2p}{(2\pi)^2}
    \frac{d^2k}{\left(2\pi\right)^2}\left|\Gamma\left(k^+;P\right)S\left(k\right)\right|^2\\
    &\times\left(2\pi\right)^2\delta^{(2)}\left(k-P-p\right)
   (2\pi) \delta\left(k^+ - P^+/z\right) \nonumber\\
  &\times \Theta(p^+)(2p^+)\left(2\pi\right)\delta\left(p^2\!-\!m^2\!+\!2\lambda\!-\!\frac{2p^+\lambda}{\rho}\right).
\label{eq:FF_feyn_diag_approach}
\end{align}

The last $\delta$-function in \eqref{eq:FF_feyn_diag_approach} comes from the cut of the quark propagator in Fig.~\ref{fig:feyn_diag_FF}. 
To understand this, we start from the identity:
\begin{align}
  S_F(p) = & \frac{2ip^+}{p^2-m^2+2\lambda-2\left|p^+\right|\frac{\lambda}{\rho}-i\mathrm{sgn}(p^+)\epsilon} \nn\\
    & + \Theta(p^+)\left(2p^+\right)\left(2\pi\right)\delta\left(
      p^2\!-\!m^2\!+\!2\lambda\!-\!\frac{2p^+\lambda}{\rho}
    \right).
\label{eq:prop_cut}
\end{align}
The first term signifies the advanced propagator, which  does not contribute to the cut, while the second term contributes to the cut.

From \eqref{eq:prop_cut}, we can derive the dispersion relation of an on-shell quark
\begin{align}
  p^- = \frac{m^2 - 2\lambda}{2p^+} + \frac{\lambda}{\rho},
\end{align}
which agrees with \eqref{eq:H0_on_quark} derived from the Hamiltonian approach.

The LF energy of the on-shell quark depends on $\rho$, which can be related to an $x^-$-independent gauge transformation. 
However, if we consider gauge-invariant quantities such as the fragmentation function, 
the dependence of $\rho$ must be eliminated in the final result.

Substituting the Feynman rules \eqref{eq:dressed_quark_prop} and \eqref{eq:Hqqbar_vertex} into \eqref{eq:FF_feyn_diag_approach}, 
and integrating out the $\delta$-functions, we obtain
\begin{align}
  D_{q\to H}\!\left(z\right)\!=\!\frac{4\lambda^2}{N_cz} \left[
    \frac{1}{M_H^2\!\!+\!\frac{m^2\!-\!2\lambda}{1/z\!-\!1}\!-\!\frac{m^2\!-\!2\lambda}{1/z}}
    \int_0^1\!\!\!dy \frac{\phi_n\left(y\right)}{\left(y\!-\!1/z\right)^2}
  \right]^2\!\!\!,
\end{align}
which exactly reproduces our master formula \eqref{eq:FF_cs} earlier derived from Hamiltonian approach. 
Hearteningly, the $\rho$ dependence arising from the initial and the final quark lines cancels out, as expected.

\end{document}